\documentclass[aps,prc,showpacs,twocolumn,superscriptaddress]{revtex4}
\usepackage{multirow}
\usepackage{graphicx}
\usepackage{amsmath}
\usepackage{mathrsfs}

\def\version{version}
\newcommand{ \be }{\begin{equation}}
\newcommand{ \ee }{\end{equation}}
\newcommand{ \bea }{\begin{eqnarray}}
\newcommand{ \eea }{\end{eqnarray}}

\usepackage{color}

\begin{document}
\title{
\begin{flushright}
{
\small \sl \version 1.3\\
%%%\today \\ for Phys. Rev. C. \\
%\Red{additions} \\
%\Green{subtractions} \\
%\Blue{modifications} \\
%\Magenta{questions} \\
}
\end{flushright}
%Correlations of event plane and effects of its decorrelation on azimuthal anisotropy measurements
Event-plane decorrelation over pseudo-rapidity and its effect on azimuthal anisotropy measurement in relativistic heavy-ion collisions
}

\author{Kai Xiao}\affiliation{Institute of Particle Physics, Central China Normal University, Wuhan, Hubei, 430079, China}
\affiliation{Key Laboratory of Quark $\&$ Lepton Physics (Central China Normal University), Ministry of Education, Wuhan, Hubei, 430079, China}

\author{Feng Liu}\email{fliu@iopp.ccnu.edu.cn}
\affiliation{Institute of Particle Physics, Central China Normal University, Wuhan, Hubei, 430079, China}
\affiliation{Key Laboratory of Quark $\&$ Lepton Physics (Central China Normal University), Ministry of Education, Wuhan, Hubei, 430079, China}

\author{Fuqiang Wang}\affiliation{Key Laboratory of Quark $\&$ Lepton Physics (Central China Normal University), Ministry of Education, Wuhan, Hubei, 430079, China}
\affiliation{Department of Physics, Purdue University, West Lafayette, Indiana 47907, USA}

\date{\today}

\begin{abstract}
Within A Multi-Phase Transport model, we investigate decorrelation of event planes over pseudo-rapidity and its effect on azimuthal anisotropy measurements in relativistic heavy-ion collisions. The decorrelation increases with increasing $\eta$ gap between particles used to reconstruct the event planes. The third harmonic event planes are found even anticorrelated between forward and backward rapidities, the source of which may root in the opposite orientation of the collision geometry triangularities. The decorrelation may call into question the anisotropic flow measurements with pseudorapidity gap designed to reduce nonflow contributions, hence the hydrodynamic properties of the quark-gluon plasma extracted from those measurements.
% event planes between the forward and backward $\eta$ ranges are surely different, especially the anti-correlation in $\Psi_{3}$ may root in the collision geometry. Besides non-flow contributions, we find event plane decorrelation will also be responsible for the reduction of azimuthal anisotropy measurements with increased $\eta$ gap.
\end{abstract}
\pacs{25.75.Ld, 25.75.Dw}

\maketitle
%--==========================================================================
\clearpage
%\section{Introduction}
%\label{sect_intro}

\textbf{\em Introduction} Relativistic heavy-ion collision data indicate that a strongly interacting quark-gluon plasma(QGP) is formed where the relevant degrees of freedom are quarks and gluons~\cite{whitepapers}. In a non-headon heavy-ion collision, the geometrical overlap region--where interactions take place between the participant nucleons--is elliptic on the transverse plane perpendicular to the collision axis. Due to interactions, the high energy density and pressure built up in the center of the collision region power an anisotropic expansion and collective motion of the QGP. This results in an elliptical distribution in the final-state particle azimuthal distribution, called elliptic flow~\cite{Ollitrault}. The measured elliptic flow is so large that hydrodynamical descriptions are applicable and the shear viscosity to entropy density ratio ($\eta/s$) cannot be much larger than the conjectured quantum low limit of $1/4\pi$~\cite{Heinz}. Similar phenomenon has been also observed in a gas of cold Fermionic lithium-6 atoms, a system very different from the QGP, where a magnetic field is used to induce strong, resonant interactions~\cite{KMO}.

It was not realized until recently~\cite{Alver} that there can be a triangular shape component in the transverse overlap region in the configuration space because of fluctuations in the nucleon distributions inside nuclei. This triangularity can result in a third harmonic in the azimuthal distribution of final particles, called triangular flow. Hydrodynamical calculations indicate that triangular flow is more sensitive to the $\eta/s$~\cite{Schenke}. Although the initial configuration space information of the overlap region is not directly observable, their footprint is contained in the final-state particle correlations. This is analogous to the nonuniform cosmological microwave background as a result of the primordial density fluctuation of the universe by gravitational interactions~\cite{Fabbri}.

\setlength{\parskip}{1ex}
Heavy-ion experiments measure anisotropic flow via final-state particle correlations. For example, one constructs an event plane (EP) to be the maximum particle emission direction, as a proxy for the participant plane--the minor symmetry axis of the nuclear overlap region in the configuration space. One then correlates a test particle with the event plane to measure anisotropic flow~\cite{Poskanzer}. As such, the measured anisotropy is contaminated by other particle correlations, generally referred to as nonflow~\cite{Borghini}. Many nonflow correlations are short ranged, such as resonance decays and jet-correlations. Thus, to reduce nonflow contributions, one often applies a pseudo-rapidity ($\eta$) gap between the particles used for EP construction and those used for anisotropy measurements. The basic assumption is that the participant plane is the same for all pseudorapidities. This may not be true as first pointed by Petersen {\em et al.}~\cite{Petersen}. This is because, unlike the reaction plane which is unique in a given event, participant planes can be different in different phase-space regions of the same event. In this study, we investigate the event plane correlations in pseudorapidity in a theoretical model and find  they are indeed different. In other words, it may not be reliable to measure anisotropic flow using the nonflow-reducing $\eta$-gap method. This finding, if also true in real data, would have important implications in terms of the QGP properties one extracts by comparing data to hydrodynamic calculations.

%\section{Analysis Method}
%\label{ampt_model}
\textbf{\em Analysis Method} We use the AMPT (A Multi-Phase Transport) model with string melting for our study.
There are four main components in AMPT:
the initial conditions, parton interactions, hadronization, and hadron interactions. The initial conditions are obtained from the HIJING model~\cite{Wang},
which includes the spatial and momentum information of minijet partons from hard processes and
strings from soft processes. The time evolution of partons is then treated
according to the ZPC parton cascade model~\cite{Zhang}. After parton interactions cease,
a combined coalescence and string fragmentation model is used for the hadronization of partons.
The scattering among the resulting hadrons is described by a relativistic transport (ART) model~\cite{Li}
which includes baryon-baryon, baryon-meson and meson-meson elastic and inelastic scatterings.

The azimuthal anisotropy is usually characterized by the Fourier coefficients~\cite{Poskanzer}:
\begin{equation}
v^{obs}_{n} = \langle\cos[n(\phi - \Psi_{n})] \rangle,
\end{equation}
where $\phi$ is the particle azimuthal angle and $\Psi_{n}$ is the n-th harmonic event plane angle. Note $\Psi_{2}$ is not necessarily the reaction plane (the plane defined by beam direction and impact parameter) due to event-by-event fluctuations.

In AMPT model $\Psi_n$ can be calculated in coordinate space by~\cite{Alver}
\begin{equation}
\Psi_{n}^{r} = \frac{1}{n}{\rm atan2}(\langle r^{2}\sin(n\phi_{part}) \rangle, \langle r^{2}\cos(n\phi_{part}) \rangle) + \frac{\pi}{n},
\end{equation}
where $r$ and $\phi_{part}$ are the polar coordinate positions of each parton.
However, the coordinate space information is not accessible by experiment. The event plane is instead constructed from measured particle momenta by
\begin{equation}
\Psi_{n}^{p} = \frac{1}{n}{\rm atan2}(\langle \sin(n\phi) \rangle, \langle \cos(n\phi) \rangle),
\end{equation}
where $\phi$ is the azimuthal angle of the particle momentum.
We therefore also study the momentum space $\Psi_{n}^p$ in AMPT.

Due to the finite multiplicity of constituents, the constructed event plane is smeared from the true one by a resolution factor. The observed anisotropy parameter needs to be corrected by the corresponding event-plane resolution as
\begin{equation}
v_{n} = \frac{v^{obs}_{n}}{\mathscr{R}_{n}}.
\end{equation}
The resolution factor can be obtained by the sub-event method with an iterative procedure~\cite{Poskanzer}. Because of the large initial parton multiplicity, our calculated $\mathscr{R}_{2}^{r}$ and $\mathscr{R}_{3}^{r}$ are nearly unity, even in the most forward or backward $\eta$ range ($3.5<|\eta|<4$). However, the resolution on the final-state momentum space event planes deviate significantly from 1. For 20-50\% centrality Au+Au collisions, $\mathscr{R}^{p}_{2}$ ($\mathscr{R}^{p}_{3}$) decreases from 0.302 (0.053) at $\eta$=0-0.5 to 0.072 (0.008) at $\eta$=3.5-4.
%one first estimates the resolution parameter $\chi_{sub}$ of subevents by solving the equation
%\begin{equation}
%\mathscr{R}_{n}(\chi_{sub}) = \sqrt{\langle \cos[n(\Psi_{n}^{a} - \Psi_{n}^{b})] \rangle},
%\end{equation}

%where $\mathscr{R}_{n}$ is defined by
%\begin{equation}
%\mathscr{R}_{n}(\chi) = \frac{\sqrt{\pi}}{2}e^{-\chi^{2}/2}\chi[I_{0}(\frac{\chi^{2}}{2})+I_{1}(\frac{\chi^{2}}{2})],
%\end{equation}
%where $I_{0}$ and $I_{1}$ are modified Bessel functions. The full event plane resolution is obtained by
%\begin{equation}
%\mathscr{R}_{n} \equiv \mathscr{R}_{n}(\chi) = \mathscr{R}_{n}(\sqrt{2}\chi_{sub})
%\end{equation}

%
%, $\mathscr{R}_{n} = \langle \cos[n(\Psi_{n} - \Psi_{RP})] \rangle$.

\begin{figure}[t]
\renewcommand{\figurename}{FIG.}
\vskip -0.5cm
\includegraphics[width=0.5\textwidth]{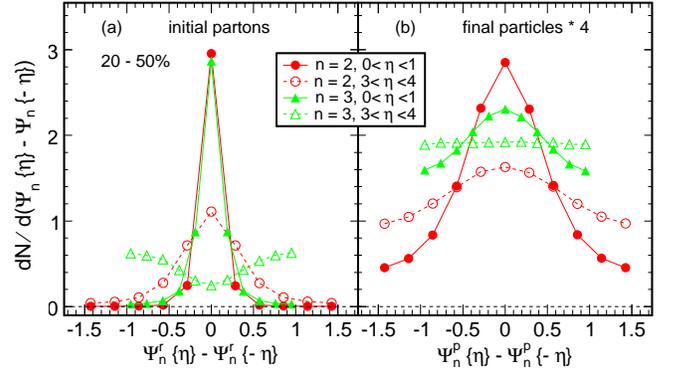}
\vspace{-3em}
\caption{(Color online)  Event probability distribution in the difference of event planes constructed by (a) initial parton coordinates and (b) final particle momenta
 in the forward and backward $\eta$ ranges for 20-50\% centrality in Au+Au collisions at $\sqrt{s_{NN}}$ = 200 GeV from the AMPT model.}
\vspace{-1em}
\label{Plot::difPsi}
\end{figure}

\begin{figure}[t]
\vskip 0cm
\includegraphics[width=0.5\textwidth]{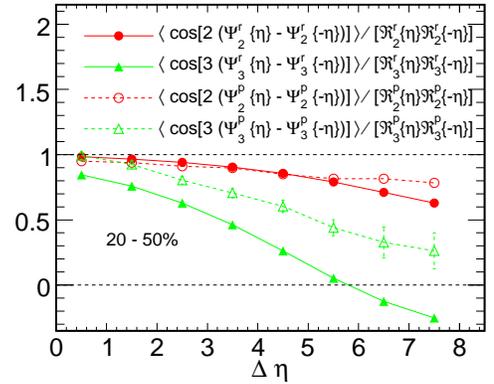}
\vspace{-3em}
\caption{ (Color online) Pseudo-rapidity gap dependence of the correlation strength $\langle\cos[n(\Psi_{n}\{\eta\}-\Psi_{n}\{-\eta\})]\rangle$ corrected by the corresponding event plane resolutions for 20-50\% centrality in Au+Au collisions at $\sqrt{s_{NN}}$ = 200 GeV from the AMPT model.}
\vspace{-1em}
\label{Plot::cosdif}
\end{figure}

\begin{figure*}[htb]
\hskip 0cm
\vskip -0.5cm
\includegraphics[width=0.95\textwidth]{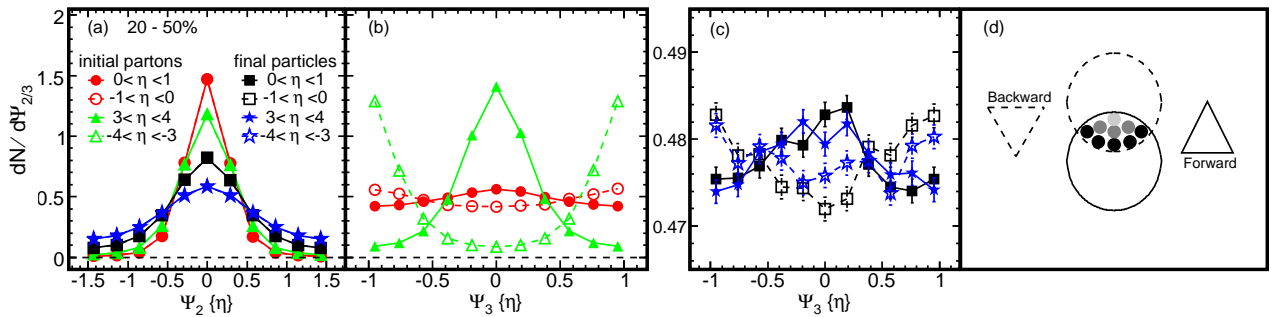}
\vspace{-1.5em}
\caption{(Color online) Event probability distribution in (a) $\Psi_{2}$ and (b),(c) $\Psi_{3}$ constructed by initial parton coordinates and final particle momenta in different $\eta$ ranges for 20-50\% centrality in Au+Au collisions at $\sqrt{s_{NN}}$  = 200 GeV from the AMPT model.  The panel (d) is a cartoon of interacting nucleons from the two nuclei.} \vspace{-1em}
\label{Plot::Psi}
\end{figure*}

%\section{Results and Discussions}
%\label{sect_discuss}

\textbf{\em Results and Discussions} Figure~\ref{Plot::difPsi}(a) shows the event probability distribution in the difference of the configuration space event-plane angles ($\Psi_{2}^{r}$ and $\Psi_{3}^{r}$) contructed from initial parton transverse coordinates in the forward and backward $\eta$ ranges. Sharp peaks are observed when $\Psi^r$ are constructed not far from midrapidity for both the elliptic and triangular harmonic planes. This indicates a strong correlation between $\Psi\{\eta\}$ and $\Psi\{-\eta\}$. With increased $\eta$ gap between the forward and backward regions, the probability distribution broadens for $\Psi_2^r$. Interestingly, the $\Psi_3^r$ from forward and background $\eta$ with large $\eta$ gap are anticorrelated on average.

The probability distribution in event-plane angle difference constructed from final-state charged particle momenta %, denoted by $\Psi_{2}^{p}$ and $\Psi_{3}^{p}$,
is shown in Fig.~\ref{Plot::difPsi}(b). Similar to the results in panel (a), the event plane angle correlations are relatively stronger close to midrapidity and become weaker when the $\eta$ gap increases. Especially, the triangular harmonic plane at very forward and backward $\eta$ are random with respect to each other. %, which indicates there is no correlation between $\Psi_{3}\{\eta\}$ and $\Psi_{3}\{-\eta\}$ in large $\eta$ range. However, the correlations of the event plane angle seem to be weaker than that of the initial spatial event plane angle, which may be due to the hadronic rescatterings.
The event-plane correlations from final-state particle momenta is weaker than those from the initial-state parton configurations. This is due to the worse resolution of the momentum space event plane, a direct result of much fewer final-state particles than the initial partons.

The decorrelation between the forward and backward event planes are, in part, due to finite event-plane resolutions. We therefore show in Fig.~\ref{Plot::cosdif} the correlation strength $\langle\cos[n(\Psi_{n}\{\eta\}-\Psi_{n}\{-\eta\})]\rangle$ divided by the corresponding event plane resolutions %$\mathscr{R}_{n}$, where the $\mathscr{R}_{n}$ is the product of $\mathscr{R}_{n}\{\eta\}$ multiplied by $\mathscr{R}_{n}\{-\eta\}$.
$\mathscr{R}_{n}\{\eta\}\times\mathscr{R}_{n}\{-\eta\}$.
The correlation is plotted as a function of the $\eta$ gap ($\Delta\eta$, defined as the difference between the centers of the $\eta$ regions used for event-plane calculation); $\Delta\eta\equiv2\eta$ for our choice of the symmetric $\eta$ ranges.
Ideally, when $\Delta\eta$ is small close to zero, $\Psi\{\eta\}$ should be the same as $\Psi\{-\eta\}$. Indeed Fig.~\ref{Plot::cosdif} shows the event-plane correlation approaches unity at small $\Delta\eta$.
However, we find that the correlation decreases with increasing $\Delta\eta$, falling siginificantly below unity.
This indicates that the event-plane decorrelation is not simply due to the degrading event-plane resolution, but physics--the event planes from forward and backward pseudorapidities are indeed different. %This effect is much stronger in $\Psi_3$ than in $\Psi_2$; with large enough $\eta$-gap, the third harmonic event planes from configuration space are anti-correlated.
The decorrelation of $\Psi_{2}^{r}$ and $\Psi_{2}^{p}$ are similar as function of $\Delta\eta$. %can see that the magnitudes of the correlation strength of $v_{2}$ event plane angles between $\Psi_{2}^{r}$ and $\Psi_{2}^{p}$ are almost consistent with each other except for the large $\eta$ gap.
The decorrelations in the triangular event planes are much stronger than those in the elliptic ones. They also have different $\Delta\eta$ dependences; The decorrelation magnitudes differ between $\Psi_{3}^{r}$ and $\Psi_{3}^{p}$. %the correlation strength shows smaller values than that of $v_{2}$ event plane angles, which also decrease faster as the $\eta$ gap increases.
With large enough $\eta$-gap, the triangular event planes from configuration space are anti-correlated.

To illustrate the physics further, we show in
Fig.~\ref{Plot::Psi}(a) the event probability distribution in $\Psi_{2}$, constructed from both the initial parton coordinates and the final particle momenta in different $\eta$ ranges, relative to the reaction plane angle (fixed at zero). %Although there are some differences in the event-by-event distribution between $\Psi_{2}$ in forward and backward $\eta$ ranges, as shown in Fig.~\ref{Plot::difPsi} and Fig.~\ref{Plot::cosdif}, we find the whole event distributions of $\Psi_{2}$ in forward and backward $\eta$ ranges are in an agreement with each other.
The forward and backward $\Psi_{2}$ distributions are consistent with each other.
The $\Psi_{2}$ distribution is wider than the $\Delta\Psi_2$ distribution in Fig.~\ref{Plot::difPsi}, even though there is decorrelation beyond event-plane resolution in $\Delta\Psi_2$. This suggests that the reconstructed event-plane does not reflect the reaction plane but the participant plane; There exists an additional fluctuation in the participant plane about the reaction plane. In other words, the two non-identical participant planes (one at forward and the other at backward pseudo-rapidity) are closely correlated, but both deviate significantly from the reaction plane.
%It is seen that the $\Psi_{2}$ distribution in the middle $\eta$ range is sharper than that in the most forward or backward $\eta$ ranges. This difference may cause some uncertainties to anisotropic flow measurements, which will be studied in Fig.~\ref{Plot::vn}.

The $\Psi_{3}$ distributions are displayed in Fig.~\ref{Plot::Psi}(b). Unlike $\Psi_{2}$, the forward and background $\Psi_{3}$ are anticorrelated, especially between those with a large $\Delta\eta$ gap. As the system expands, the property of this anti-correlation in the initial geometry is transfered to the final state particles in the momentum space, although at a much reduced magnitude. This is shown in Fig.~\ref{Plot::Psi}(c).%This is shown in the zoomed-in insert in Fig.~\ref{Plot::Psi}(b). [shall we make a separate panel for this plot so the figure becomes 4-panel?]% we can still see an anti-correlation between $\Psi_{3}\{\eta\}$ and $\Psi_{3}\{-\eta\}$.

The anti-correlation in $\Psi_3$ may root in the collision geometry. This is illustrated in the cartoon of Fig.~\ref{Plot::Psi}(d). The nucleons in the overlap region of the forward-going projectile nucleus (indicated by the solid sphere) see various thicknesses of the backward-going target nucleus (indicated by the dashed sphere). The projectile nucleons indicated by the dark dots suffer fewer collisions than those indicated by the lighter ones, thus have a relatively larger probability to end up in the forward direction. In addition there are more of the ``dark'' nucleons than the light ones in the projectile nucleus. Therefore, at forward rapidity, the distribution of the struck nucleons is more likely to have a upward triangular shape (indicated by the solid triangle). Conversely, that at backward rapidity is more likely shaped by the oppositely oriented dashed triangle. Thus, triangularities of the forward and backward interaction zones are more likely anticorrelated.

\begin{figure}
\vskip -1cm
\includegraphics[width=0.5\textwidth]{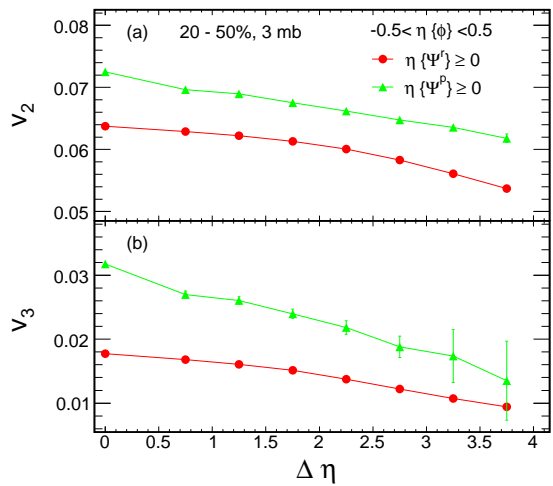}
\vspace{-2.5em}
\caption{(Color online) Pseudo-rapidity gap dependence of (a) $v_{2}$ and (b) $v_{3}$ with respect to $\Psi^{r}$ and $\Psi^{p}$ from different $\eta$ ranges with 3 mb parton cross section for 20-50\% centrality in Au+Au collisions at $\sqrt{s_{NN}}$ = 200 GeV from the AMPT model.} \label{Plot::vn}
\vspace{-1em}
\end{figure}

Experimentally, to reduce nonflow, one often measures particle anisotropy by applying a $\eta$-gap between the particles of interest and those used for event plane reconstruction. The event-plane decorrelation we have shown calls this $\eta$-gap method into question. This was first pointed out by Petersen {\em et al.}~\cite{Petersen}. Here we study the effect of event-plane decorrelation on anisotropy measurements. In Fig.~\ref{Plot::vn}, we investigate $v_{2}$ and $v_{3}$ of charged particles within $|\eta|<0.5$ with respect to $\Psi^r$ and $\Psi^p$ from different $\eta$ ranges as functions of the $\eta$ gap. The $v_n$ is already corrected by the corresponding event-plane resolutions. As the $\eta$ gap increases, the magnitudes of $v_{2}$ and $v_{3}$ decrease. The decrease is, in part, due to the decorrelation between the event plane at forward $\eta$ and the event plane at $|\eta|<0.5$. The decorrelation increases with increasing $\Delta\eta$. This study suggests that one may be required to measure particle anisotropy using the event plane reconstructed from the same $\eta$ region as the particles of interest.

The decreasing behavior of $v_n$ with $\Delta\eta$ is found for both $\Psi_n^r$ and $\Psi_n^p$. However, the interpretation is complicated for $v_n$ using $\Psi_n^p$ due to nonflow. Nonflow is generally also a decreasing function of $\Delta\eta$. The decreasing trends of $v_{n}$ with respect to $\Psi_n^{p}$ is therefore a result of combined effects from the reduced non-flow and decorrelation in the event plane angles over $\eta$ gap.

%In fact, the difference between the two $v_2$ data sets in Fig.~\ref{Plot::vn}(a) is due to nonflow because the event-plane decorrelations with $\Delta\eta$ are similar for $\Psi_2^r$ and $\Psi_2^p$ (see Fig.~\ref{Plot::cosdif}). The difference between the two $v_3$ data sets in Fig.~\ref{Plot::vn}(b) is not caused only by nonflow because the event-plane decorrelations are different for $\Psi_3^r$ and $\Psi_3^p$ (see Fig.~\ref{Plot::cosdif}).
%Since non-flow is sensitive to the chosen $\eta$ gap, larger $\eta$ gap will lead to smaller non-flow. Based on our study, we find that the event plane angle is not constant over pseudo-rapidity, which will cause some uncertainties to the $v_{2}$ and $v_{3}$ measurements. Therefore,

Is the difference in $v_{n}$ from $\Psi_n^{r}$ and from $\Psi_n^{p}$ entirely due to nonflow? The answer is no, because the event plane decorrelations are different for $\Psi_3^{r}$ and from $\Psi_3^{p}$, which must result in a correlation $\langle \cos[3(\Psi_3^{r}\{\eta\} -\Psi_3^{p}\{\eta\})] \rangle$ varying with $\eta$. Even with the similar decorrelations for $\Psi_2^{r}$ and $\Psi_2^{p}$ as a function of $\eta$, we found that $\langle \cos[2(\Psi_2^{r}\{\eta\} -\Psi_2^{p}\{\eta\})] \rangle$ varies with $\eta$. In other words, the difference in $v_n$ between $\Psi_n^r$ and $\Psi_n^p$ is a combined effect of nonflow and $\langle \cos[n(\Psi_n^{r}\{\eta\} -\Psi_n^{p}\{\eta\})] \rangle$. The fact that $\Psi_n^r\{\eta\}$ and $\Psi_n^p\{\eta\}$, after resolution corrections, are not the same may suggest that the final state particles in a given $\eta$ region are not solely determined by the initial partons from the same $\eta$ region. This is in fact not completely unexpected but the discussion of its physics is beyond the scope of this paper.

\textbf{\em Conclusions} By utilizing AMPT model with string melting, we have studied the correlations between event plane azimuthal angles reconstructed from particles in the forward and backward peudo-rapidity regions in Au+Au collisions at $\sqrt{s_{NN}} = 200 GeV$. The event planes recontructed from the initial parton coordinate space and the final particle momentum space are both studied. The event plane correlations are corrected by the event-plane resolutions. The resolution corrected event-plane correlation is found to weaken as the forward-backward $\eta$-gap increases. This indicates that the event planes from different $\eta$ ranges differ. The decorrelation in the triangular event plane is significantly stronger than that in the elliptic one. Particularly, the initial parton coordinate $\Psi_{3}$ exhibits an anti-correlation between very forward and backward $\eta$. The anticorrelation may root in the collision geometry--the triangularity of the forward going participants is opposite to that of the backward going ones.

The decorrelation may call into question the anisotropic flow measurements with pseudorapidity gap designed to reduce nonflow contributions.
Because of the event-plane decorrelation, azimuthal anisotropy measured by the event planes from different $\eta$ ranges yield different results. By applying a large $\eta$ gap between the anisotropy measurements and the event-plane recontruction, one can reduce the nonflow contributions, but at the same time increase the event-plane decorrelation effect. It may therefore not be a reliable strategy to apply a large $\eta$ gap to measure anisotropic flow. The large $\eta$-gap may result in an under-measured anisotropic flow because of the event-plane decorrelation. If this were true in real data, it could have important implications to the QGP properties extracted by comparing data to hydrodynamic calculations. In particular, one may speculate that the extracted shear viscosity to entropy density ratio could be overestimated due to an under-measured flow, which in turn could suggest that the QGP may be more perfect as we thought it was.

%\vspace{0cm}

%\section{Acknowledgments}
\textbf{\em Acknowledgments} This work is supported in part by the National Natural Science Foundation of China under grant No. 11228513, 11075060, 11135011 and U.S. Department of Energy under Grant No. DE-FG02-88ER40412 and CCNU-QLPL Innovation Fund (QLPL2011P01).

%
%\clearpage
%--=================================================================

%
%%--==========================================================================
\end{document}